\documentclass[aps,pre,twocolumn,groupedaddress,showpacs,floatfix]{revtex4}

\usepackage[dvips]{graphicx}
\usepackage{amsmath}
\usepackage{amssymb}

\begin{document}


\title{The Ising model in a Bak-Tang-Wiesenfeld sandpile}

\author{Zbigniew Koza} 
\affiliation{Institute of Theoretical Physics, University of Wroc{\l}aw,
pl.\ M.\ Borna 9, 50204 Wroc{\l}aw, Poland}

\author{Marcel Ausloos}
\affiliation{
SUPRATECS, B5, Li\`ege, Belgium}

\date{\today}

\begin{abstract}
  We study the spin-1 Ising model with non-local constraints imposed by the
  Bak-Tang-Wiesenfeld sandpile model of self-organized criticality (SOC).  The
  model is constructed as if the sandpile is being built on a {(honeycomb)}
  lattice with Ising interactions.  {In this way we combine two
    models that exhibit power-law decay of correlation functions characterized
    by different exponents.}  We discuss the model properties through an order
  parameter and the mean energy per node, as well as the temperature
  dependence of their fourth-order {Binder} cumulants. We find (i) a
  thermodynamic phase transition at a finite $T_c$ {between paramagnetic
    and antiferromagnetic phases}, and (ii) that above $T_c$ the correlation
  functions decay in a way typical of SOC.  The usual thermodynamic
  criticality of the two-dimensional Ising model is not affected by SOC
  constraints {(the specific heat critical exponent $\alpha \approx 0$)},
  nor are SOC-induced correlations affected by the interactions of the Ising
  model.  Even though the constraints imposed by the SOC model induce
  long-range correlations, as if at {standard} (thermodynamic) criticality,
  these SOC-induced correlations have no impact on the thermodynamic
  functions.
\end{abstract}


\maketitle

\section{Introduction}
\label{sec:Introduction}

The phenomenon of self-organized criticality (SOC) attracts a lot of interest
in various branches of science (see \cite{JensenBook,Dickman00} for reviews).
Its most intriguing feature resides in the ``spontaneous'' formation of
scale-free spatio-temporal patterns with power-law correlations between
various quantities.  On the one hand, these correlations closely resemble
those that appear at critical points in continuous phase transitions. On the
other hand, while the critical state in phase transitions is temperature- or
external-field-driven, a SOC state is {thought}  to form
without fine-tuning of any external parameters.  The relation between the
``classical'' and ``self-organized'' criticality has been studied by many authors
\cite{Dickman00,Sor95,Vesp98,Meester04u} with a conclusion that the origin of
the self-organized criticality is a continuous absorbing-state phase
transition to which a SOC system is attracted.

Perhaps the best-known and most extensively studied model of thermodynamic
criticality is the Ising model of ferromagnetism.  Of similar importance for
self-organized criticality is the Bak-Tang-Wiesenfeld (BTW) model \cite{Bak88}
of a sandpile. Owing to its elegant mathematical structure and many rigorous
results \cite{DharAuto90,Dhar99,Meester01,Ivash98}, the BTW model has
practically become a paradigm for self-organized criticality studies.

Specific SOC features of the BTW model, which will attract our attention,
include the fact that in {the BTW} model correlation functions decay with
the distance $r$ as $r^{-2d}$, where $d$ is the space dimensionality
\cite{Majumdar91,Ivash94}, {which resembles, but differs from the
  $r^{-(d-2+\eta)}$ decay in standard critical phenomena.} As the behavior of
a correlation function is a key ingredient of criticality, we will use this
property to build and analyze a model which, by construction, is expected to
exhibit {long-range correlations induced by} both classical and
self-organized critical phenomena.  In defining such a model we follow the
idea of Ref.~\cite{Dinaburg04}: {take a standard model of statistical
  physics and significantly reduce its phase space to that of a corresponding
  SOC model.}

One of the main results of ref.\ \cite{Dinaburg04}, where a combination of the
Potts and BTW models was investigated, is that in the limit of a vanishingly
small temperature their hybrid system looses its `self-organized criticality',
i.e.\ the power-law correlations disappear.  {Thus it seems worthwile to}
investigate a hybrid model, numerically, for a wide range of temperatures. In
particular, we want to check if the self-organized criticality imposed by
nonlocal constraints can be detected in the behavior of thermodynamic
functions. However, as it was pointed out in Ref.~\cite{Dinaburg04},
{because of the non-local nature of the applied constraints}, analysis of
this type of models is a difficult problem, still at its infancy, so it is of
interest to check whether standard concepts and theorems of statistical
physics, like existence of the thermodynamic limit, ergodicity,
fluctuation-dissipation theorem or the central limit theorem apply to
such hybrid systems.

In our study we choose the two-dimensional spin-1 Ising model as
the Hamiltonian system and the BTW sandpile model as the SOC component.  The
resulting model is an equilibrium model that takes (short-range) interactions from the Ising model and
the (non-local) constraints from the BTW model.  Note, however, that while
the Ising model focuses on spin configuration at thermodynamical
equilibrium, the BTW model attempts to describe a nonequilibrium, dynamical
process where grains of sand are continually added to the lattice to form
``avalanches'' of different sizes and durations.

The construction of such a ``hybrid'' model employs the fact that both
component models are built on a lattice with special variables (``spins''
$s_j$ or ``grain heights'' $h_j$) attached to each lattice node, and the
values of these variables are limited to a few integers, e.g.\ $s_j
\in\{-1,0,1\}$ or $h_j\in\{0,1,2\}$. This property allows us to map the states
of one model onto the states of the other one.  
{To reduce to the minimum
the number of possible mappings of the spins $s_j$ of the Ising model to the
sand heights $h_j$ of the BTW sandpile, we choose a two-dimensional (2D)
honeycomb lattice (coordination number $z=3$). }

In a hybrid system like this the high-temperature limit is determined entirely
by the constraints of the SOC component; consequently, the spin correlation
function should decay with the distance $r$ as $r^{-4}$
\cite{Majumdar91,Ivash94}. On the other hand, the ground state is expected to
be controlled mainly by the Hamiltonian, and at sufficiently low temperatures
interactions should force the system into an ordered phase.  Consequently, one
can expect a thermodynamic phase transition at some finite temperature $T_c$.
For the standard 2D Ising model without an external field this transition is
continuous {and the spatial correlations decay as $r^{-\eta}$ with $\eta =
  1/4$ \cite{Fisher}}.  Therefore, if introduction of the SOC constraints turns out to be
too weak to change the nature of this transition, the hybrid model should
exhibit a power-law decay of correlation functions both for $T\to\infty$ and
at $T_c$.

Although the Ising model has been modified in many, many ways, most of the
modifications are of local character. These include building the model on
fractal {\cite{Gefen80} or small-world  \cite{SmallWorld}} lattices as well
as introduction of quenched \cite{Glass:Binder,Glass:Jackle} or kinetic
\cite{Majumdar02,Glass:Newman99} disorder. In our present approach the
modification is nonlocal -- one has to scan the whole lattice to decide
whether a given configuration is allowed or not. This introduces several
complications, but at the same time opens new directions of research. 
In our
analysis we shall consider both the effects of introducing interactions and
temperature to a SOC model and the changes introduced by a SOC-like
constraints into a Hamiltonian system.

The paper is organized as follows. In section \ref{sec:Model} we define the
model. Then, in section \ref{sec:Numerics} we discuss problems that appear in
Monte Carlo simulations and we present some methods that we have applied to
overcome them.  The results are presented in section \ref{sec:Results}.  They
include a numerical analysis of {the finite-size effects imposed by the SOC
  subsystem}, an analytical study of the influence of the sandpile model on
the number of phases in the Hamiltonian subsystem, and a numerical analysis of
the impact of the SOC and Hamiltonian components on the critical properties of
the system.  It also contains results of several tests carried out to verify
if the hybrid model {with nonlocal constraints} can be treated with
standard methods of statistical physics. These include tests of whether
  fluctuations of the internal energy can be used to determine the specific
  heat and whether far from the critical point the central limit theorem can
  be applied to predict the distribution of the energy fluctuations.
Finally, section \ref{sec:conclusions} is devoted to conclusions.

\section{Model}
\label{sec:Model}

\begin{figure}
\includegraphics[width=0.45\columnwidth, clip=true]{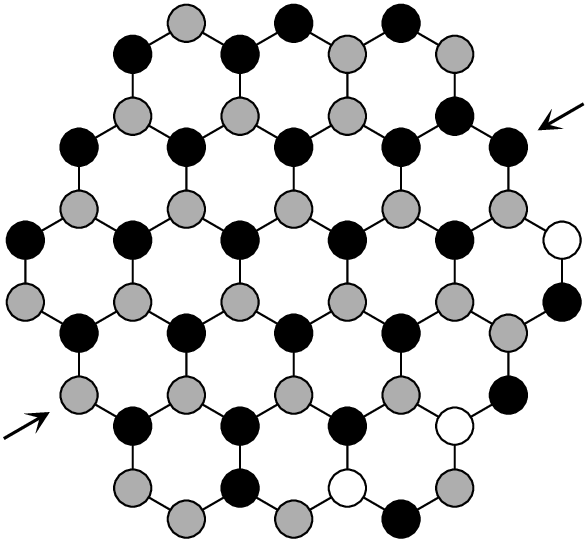}
    \caption{An example of an allowed {BTW} configuration on a honeycomb lattice of
    linear size $L=3$. Empty, shadowed, and filled circles represent the nodes
    with heights $h_j = 0, 1, 2$ (or spins $s_j = 0,-1,+1$), respectively. The
    arrows show an example of a pair of boundary nodes that interact via
    {the Ising} Hamiltonian (\protect\ref{eq:Ham}).
    \label{fig:1}
    }
\end{figure}

We define the model on a two-dimensional honeycomb lattice of linear size $L$.
An example of such a lattice with $L=3$ is shown in Fig.~\ref{fig:1}. The
lattice has a shape of a big hexagon made of $3L(L-1)+1$ small hexagons and it
has $N=6L^2$ nodes. Most of them are interior nodes connected to 3 nearest
neighbors, but $6L$ nodes lie on the edge of the lattice and have only 2
connections.

Each node $j$ can be in one of three states, which can be interpreted either as
a spin variable $s_j\in\{-1,0,1\}$ or the number of sand grains
$h_j\in\{0,1,2\}$. Of 6 mappings of heights $h_j$ on spins $s_j$, we choose
the one in which $h_j = 0,1,2$ corresponds to $s_j=0,-1,+1$, respectively
{(actually, due to the $Z_2$ symmetry, the number of non-equivalent
  mappings reduces to 3)}.

The Hamiltonian of the model is simply that of the spin-1 Ising model,
 \begin{equation}
   \label{eq:Ham}
    H = -J \sum_{\left<i,j\right>} s_i s_j
 \end{equation}
where $J$ is a real parameter, $s_j = 0,\pm1$, and the sum is to be taken over
all pairs of nearest neighbors.
To minimize finite-size effects, we have adopted periodic
boundary conditions. To this end we assume that, as is illustrated in
Fig.~\ref{fig:1}, each boundary node interacts, via (\ref{eq:Ham}), not only
with its two direct lattice neighbors, but also with the node at the location
symmetric about the lattice center. 
This choice ensures that the ground state
in the antiferromagnetic case ($J<0$) is doubly degenerated and, together with 
the several consecutive excited states,  
obey the  $Z_2$ symmetry.

In the standard spin-1 Ising model the phase space is defined by all possible
spin configurations. To each such configuration $\eta$ one can assign a
non-vanishing temperature-dependent probability $\propto \exp[-H(\eta)/k_BT]$,
and their number is exactly $3^N$. However, the states of the BTW model, which
is a dynamic, nonequilibrium model of a sand pile in a continual flux of sand
grains, have completely different properties. It turns out that 
all configurations can be divided into two categories: disallowed and
allowed. The former can never be found in the steady state, while the latter
can be found with the same probability \cite{DharAuto90,Dhar99,Meester01}.
Moreover, the number of allowed configurations is $a^N$ with $a<3$
\cite{Dhar99}; hence the number of allowed configurations is only a fraction 
of all configurations and this fraction tends to 0 as $L\to\infty$.

The main idea behind building a hybrid Hamiltonian-SOC model is to take  a
standard model of equilibrium statistical physics, e.g.\ the spin-1 Ising model, and
significantly reduce its phase space to that of a corresponding SOC model,
e.g.\ the BTW model. Solving a hybrid model is then equivalent to finding the
partition function
\begin{equation}
 \label{eq:pf}
   Z = \sum_{\eta} \Theta(\eta) \exp[-H(\eta)/k_B T],
\end{equation}
where the sum runs over all states $\eta$ of the Hamiltonian component,
$\Theta$ is the characteristic function of a SOC state,
\begin{eqnarray*}
  \Theta(\eta) &=& 1,\quad \mbox{if } \eta \mbox{ is allowed}  \\
  \Theta(\eta) &=& 0,\quad \mbox{if } \eta \mbox{ is disallowed}
\end{eqnarray*}
and $k_B$ is the Boltzmann constant. 

To complete the definition of the model, it remains to explain how to
distinguish an allowed configuration from a disallowed one in the BTW model. A
complete discussion of this problem is given by Dhar in
\cite{DharAuto90,Dhar99}; here we will give only a brief summary 
{of his results:}  (i) a configuration with all $h_j=2$ is allowed. (ii) If we take
an allowed configuration $\{h_j\}$ and increase the value of $h_j$ at some $j$
by one, this $h_j$ may exceed the maximum value $2$; however, if we then relax
such an ``unstable'' state (in the way defined below), we will always arrive at
an allowed configuration. (iii) All allowed configurations can be reached by
iteratively applying step (ii) to any allowed configuration. Relaxation of
unstable states is carried out as follows: for any site $j$ at which $h_j > 2
$, replace $h_j$ with $h_j - 3$ and for every neighbor $k$ of $j$ increase
$h_k$ by 1.  Relaxation of any state can be carried out in arbitrary order,
consists of a finite number of steps that form the so called avalanches and
the resulting stable state is unique \cite{DharAuto90,Dhar99}. An example of
an allowed configuration is shown in Fig.~\ref{fig:1}. Note that while we use
periodic boundary conditions for the Hamiltonian part of the model, we employ
open BC for the relaxation process so that excessive ``sand grains'' can leave
the system.

\section{Numerical implementation}
\label{sec:Numerics}

Most sandpile models studied so far are characterized by irreversible
dynamics, where one only \emph{adds} grains which later leave the system
through open boundaries.  Thus, if by adding a grain to a stable state $X$ and
relaxing it one arrives at another stable state $Y$, there is usually no way
to return from $Y$ to $X$ in just one step.  As this violates the
detailed-balance condition and hence drives the system out of equilibrium, in
our simulations we also use the method of Ref.~\cite{Dhar94} to construct the
reversed process that transforms $Y$ back to $X$.  This method is based on
removing a grain from an (arbitrary) node and then fully relaxing the
resulting state.
 
 Following the standard
Metropolis algorithm \cite{Metropolis,Binder97}, in each time step we
construct a trial state by picking at random a lattice node $j$ and adding or
removing a grain at it.  If this renders the configuration unstable, 
an avalanche is generated and the system is fully relaxed to a stable state.
We then calculate the energy of the trial state,
$E'$, compare it with the energy of the original state, $E$, and accept the
trial state  with probability $\min(1,\exp[-(E'-E)/k_B T])$. This, however, may
lead to problems with ergodicity.  Since the phase space of our model is
limited to recurrent states of a Markov process defined by the BTW model, there
is no doubt that at least in theory the system is ergodic. However, some
transition probabilities can be extremely small.  This problem can be
particularly serious for transitions related to large avalanches, as in their
case the factor $\exp[-(E'-E)/k_BT]$ can quickly tend to 0 as $L\to\infty$.

Calculation of a trial state in our model is extremely time-consuming, for it
requires to perform full relaxation of an excited state, and the average
number of individual topplings in an avalanche is proportional to the number
of lattice nodes $N$ \cite{DharAuto90}.  For this reason we were able to
perform simulations only for rather small lattices with $N \le 9600$ (for a
pure Ising or BTW models this number could be easily increased by a factor
1000).  Due to the non-local nature of sandpile dynamics employed to generate
trial states, there seems to be no way to apply here any of the many
techniques for speeding up Monte-Carlo simulations \cite{Binder97}. The only
`trick' we used was to store, for a given configuration, energies of any
rejected Monte-Carlo trial steps to avoid multiple relaxations of the same
state. For low temperatures this can speed up calculations by a factor of 100.

\section{Results}
\label{sec:Results}

\subsection{{Finite-size effects induced by the BTW component}}
\label{sub:T2infty}

{Two essentially different types of finite size effects can be expected to
  appear in the model.  The first kind, typical of critical points, results
  from the fact that at criticality the diverging correlation lengths exceed
  the system size.  The second kind is inherent to 
  SOC models with open boundaries, as the mean concentration of particles near
  the open boundaries tends to be smaller than that in the bulk. For a
  two-dimensional lattice of linear size $L$ this phenomenon brings about a
  finite-size correction of order $O(1/L)$ \cite{Grass90}.  This
    finite-size effect has a noticeable impact on the 
    thermodynamics of the hybrid model \emph{at all temperatures}}. Its magnitude
  can be appreciated by studying the limit of the temperature going to
  infinity, since in this case our model is governed entirely by its SOC
  component.

\begin{figure}
\includegraphics[width=\columnwidth, clip=true]{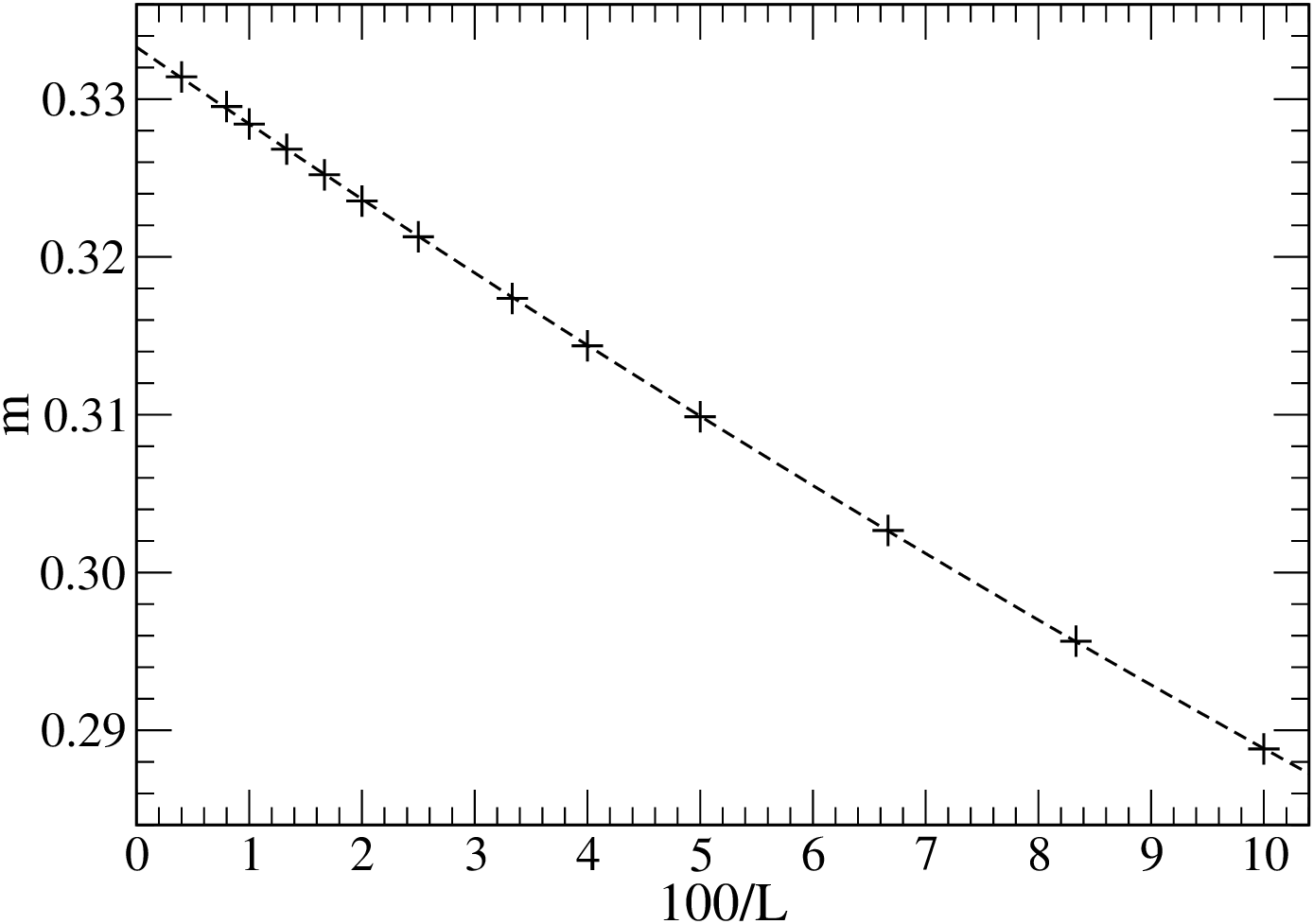}
    \caption{Order parameter $m$ as a function of $L^{-1}$ for
     lattice sizes $L=10,\ldots,250$. The error bars are less than the line width.
     The dashed line represents a quadratic fit and the extrapolated value for
     $L\to\infty$ is $0.33330(4)$.
    \label{fig:2}
    }
\end{figure}

{Although the finite-size effects in the BTW model were investigated in
  several papers \cite{Grass90,Lubeck97}, none of them referred to the
  honeycomb lattice.  Figure~\ref{fig:2} depicts $m = \left< s_j\right>$,
  i.e.\ the order parameter in the standard Ising model (magnetization per
  node), for several values of the system size $L$. As expected, the
  dependence of this parameter on $L$ is quite strong.}
Upon approximating $m(L)$ as a quadratic in $1/L$ and extrapolating it to
the limit of $L\to\infty$, we obtained $m_{\infty} \approx 0.33330(4)$, which
suggests that $m_\infty = 1/3$. Note that in our model $m>0$ even at infinite
temperature and in absence of an external magnetic field.  This reflects a
property of the BTW sandpile model that nodes with 2 grains (corresponding to
$h_j=2$, or $s_j=+1$) are more probable than those with 1 grain ($h_j=1$, or
$s_j=-1$).

Similarly good quadratic fits were found for {the} mean concentrations
$c^{(0)}_L$, $c^{(1)}_L$, and $c^{(2)}_L$ of nodes with $h_j = 0$, 1 and 2,
respectively. In the limit of $L\to\infty$ we obtained $c^{(0)}_\infty \approx
0.08334(5)$, $c^{(1)}_\infty \approx 0.29166(2)$, and $c^{(2)}_\infty \approx
0.62500(3)$. This suggests that $c^{(0)}_\infty = 2/24$, $c^{(1)}_\infty =
7/24$, and $c^{(2)}_\infty = 15/24$.  Note that the exact values of
$c^{(k)}_\infty$ are known for square \cite{Priezz94} and Bethe \cite{Grass90}
lattices, and numerical results are available for hypercubic lattices of
dimension 2 to 5 \cite{Grass90}.


\subsection{Effect of SOC constraints on possible phases of the Hamiltonian
system}
\label{sub:constraints}

As we have just seen, the high-temperature properties of our model are very
unusual: while in standard Hamiltonian systems at high temperatures
the magnetization disappears (the system is in a disordered, paramagnetic phase),
our hybrid system exhibits a nonzero, positive $m$ as
$T\to\infty$.  Moreover, we will show that $m\ge0$ in the limit $L\to\infty$ (of an
infinite system), at any temperature and for any values of the control
parameter $J$.

Let $N_k$ denote the total number of
lattice nodes with $k$ grains, $k=0,1,2$. On the one hand their sum is the
total number of lattice nodes,
\begin{equation}
    N_2 + N_1 + N_0 = 6L^2.
    \label{eq:N0+N1+N2=N}
\end{equation}
On the other hand, in any allowed configuration these numbers satisfy
\begin{equation}
    2N_2 + N_1 \ge B_L,
    \label{eq:2N2+N1=B}
\end{equation}
where $B_L$ is the number of bonds on a lattice of size $L$.  This relation
can be justified using the original version of the burning algorithm
\cite{DharAuto90,Dhar99}, which guarantees that the $B_L$ bonds can be mapped
onto the $N$ nodes in such a way that a node with $k$ grains is an image of at
least $k$ bonds.

Equation (\ref{eq:N0+N1+N2=N}) implies $c_L^{(2)} + c_L^{(1)} \le 1$ and,
since in our case $B_L = 9L^2 - 3L$, equation (\ref{eq:2N2+N1=B}) leads to
$2c_L^{(2)} + c_L^{(1)} \ge 3/2 - 1/2L$. Hence
\begin{equation}
    m_L = c_L^{(2)} - c_L^{(1)} = 2(2c_L^{(2)}+c_L^{(1)}) - 3(c_L^{(2)}+c_L^{(1)}) \ge -1/L.
    \label{eq:N2>N1}
\end{equation}
As the right-hand-side of this relation tends to 0 as $L\to\infty$, we conclude
that there can be no phase with a negative value of $m$. 
This is not a trivial statement, as {for suitably chosen values of $J$ and
  $T$, $m$ can take on any value in the range $[0,1]$, and} for finite
lattices there {even} exist allowed configurations with $m<0$. An example
of a system {with $m<0$} is a lattice with $L=1$ in which one node has a
hight $h_j=2$ (corresponding to the spin $s_j = +1$) and the remaining 5 nodes
have $h_j=1$ (i.e.\ $s_j = -1$).  {S}ince (\ref{eq:N2>N1}) applies to
systems at an arbitrary temperature, it implies that no
paramagnetic-ferromagnetic phase transition {at finite $T$} is possible in
the hybrid model, so that the parameter $m$ {cannot be  regarded as
a usual} magnetization.

Similar arguments lead to $c^{(0)}_L \le 1/4 + 1/4L$, which implies that no
quadrupolar phase is possible in the model (in a quadrupolar phase half of the
nodes have spins $s_j=0$, i.e.\ $c_{L}^0 \ge 1/2$). Consequently, only two
phases can exist: a high-T paramagnetic ($P$) and a low-T antiferromagnetic
($A$) phase. Note that this reduction of the number of possible phases is a
consequence of the constraints imposed on the Ising model by the sandpile
dynamics, and has nothing to do with the ``criticality'' of the latter.

{These conclusions are in accord with symmetry considerations.  Because of
  the constraints imposed by the underlying SOC model, the hybrid model has a
  single, unique ``ferromagnetic'' ground state, and hence the low-$T$ system
  with ferromagnetic coupling ($J>0$) is not invariant under the $Z_2$
  symmetry of the standard Ising model; consequently, no ferromagnetic-like
  phase transition is expected in the system. On the other hand, the SOC
  constraints do not affect the antiferromagnetic ($J <0$) ground state nor
  none of its ``neighboring'' (i.e. slightly excited) states.  Thus, the
  hybrid model with antiferromagnetic coupling at low $T$ obeys the
  antiferromagnetic $Z_2$ symmetry, which opens a possibility of the
  antiferromagnetic phase transition.  }

\subsection{Transition between paramagnetic and antiferromagnetic phases}
\label{sub:ferro-aferro}

Since we expect the antiferromagnetic phase to appear only for $J<0$, we
decided to look for a temperature-driven transition between the $P$ and $A$
phases only for $J<0$. For the sake of simplicity, we measure energy in units
of $|J|$ and set the Boltzmann constant $k_B = 1$.  To investigate a phase
transition, we formally divide the lattice into two sublattices (such that no
adjacent nodes belong to the same sublattice) and we define a parameter $a$ as
half of the difference between magnetizations on each sublattice. In the
ground state our system has a perfect antiferromagnetic order, with one
sublattice completely filled with spins $s = +1$, and the other one with
$s=-1$.  Hence the mean energy per node $u = -1.5$, magnetization $m=0$ and
the order parameter $a=\pm 1$. We expect that $|a|>0$ in the $A$ phase, and
$a=0$ in the high-$T$ phase.

We start each simulation from the ground state in phase $A_+$ with $a=1$, and
then slowly warm the system up. After equilibrating, at each temperature we
use an algorithm of \cite{Grass90} to generate $10^8$ configurations. The
results are averaged over $10$ independent runs.

\subsubsection{Internal energy and specific heat}
\label{subsub:u}

\begin{figure}
\includegraphics[width=\columnwidth, clip=true]{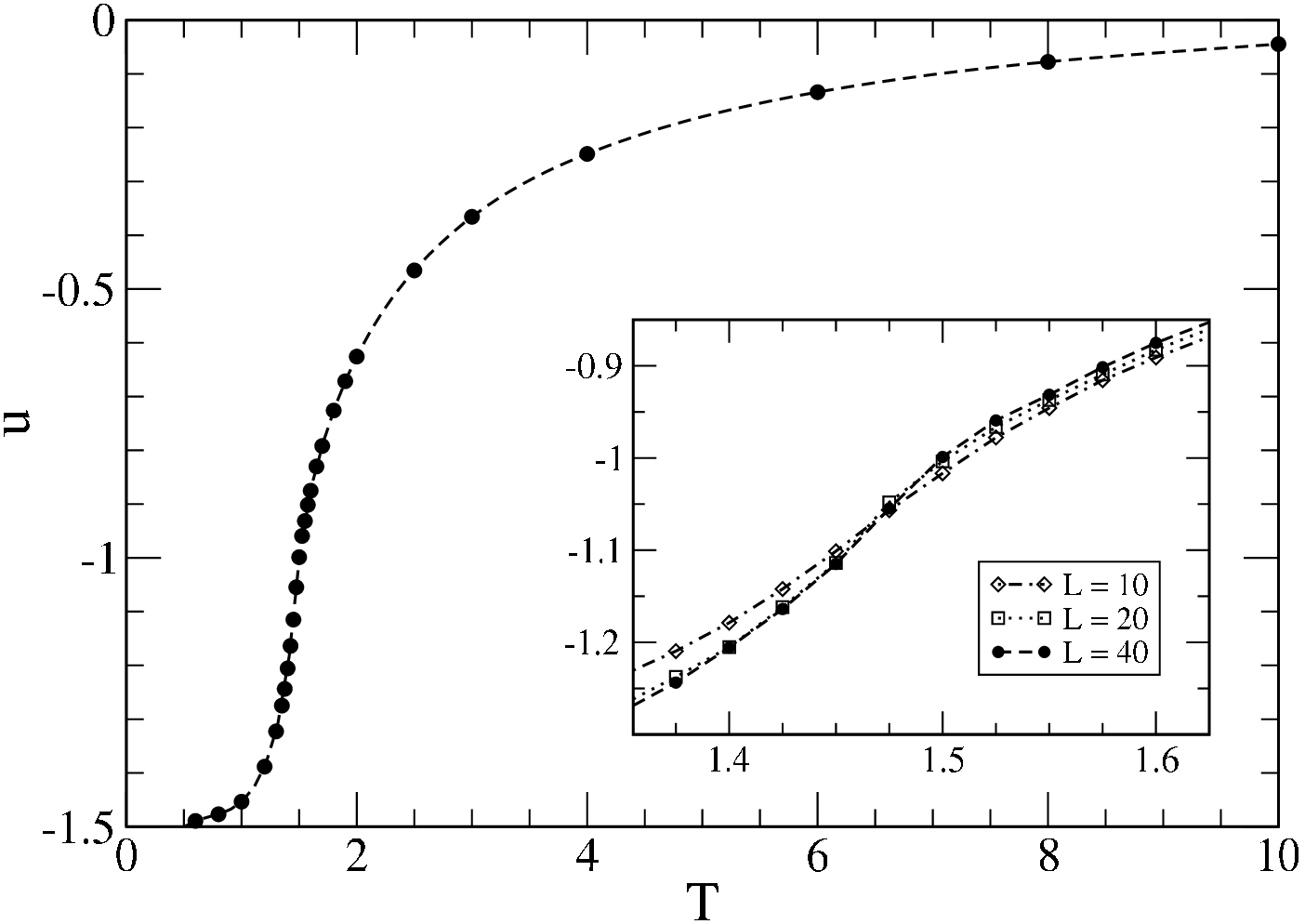}
    \caption{Energy per lattice node, $u$, as a function of temperature for
    $J<0$ and lattice size $L=40$. The inset presents the
    results for $L = 40$ ($\bullet$),  $L=20$ ($\square$), and $L=10$
    ($\diamond$) in the vicinity of $T\approx 1.475$.
    \label{fig:fig_energy}
    }
\end{figure}

In Fig.~\ref{fig:fig_energy} we present the temperature dependence of the
internal energy per lattice node, $u$, calculated for a system of linear size
$L=40$. As can be seen, the curve has a quite large slope at $T\approx1.5$. This
suggests that the specific heat
\begin{equation}
    C = du/dT.
    \label{eq:C=du/dT}
\end{equation}
may diverge near $T=1.5$, which is the first hint for a phase transition. A
blow-up of this region is presented in the inset, together with results for
smaller lattices ($L=10$ and $20$). It shows that the curves obtained for
different lattice sizes cross each other at $T_c\approx 1.475$ and suggests
that near this temperature $du/dT$ (i.e.\ the specific heat) significantly
depends on the system size. Note that this value is different from the
critical temperature of the standard spin-1 Ising model on a honeycomb
lattice, $T_c^\mathrm{Ising}\approx1.2$ \cite{Ising1,Ising2}.

\begin{figure}
\includegraphics[width=\columnwidth,clip=true]{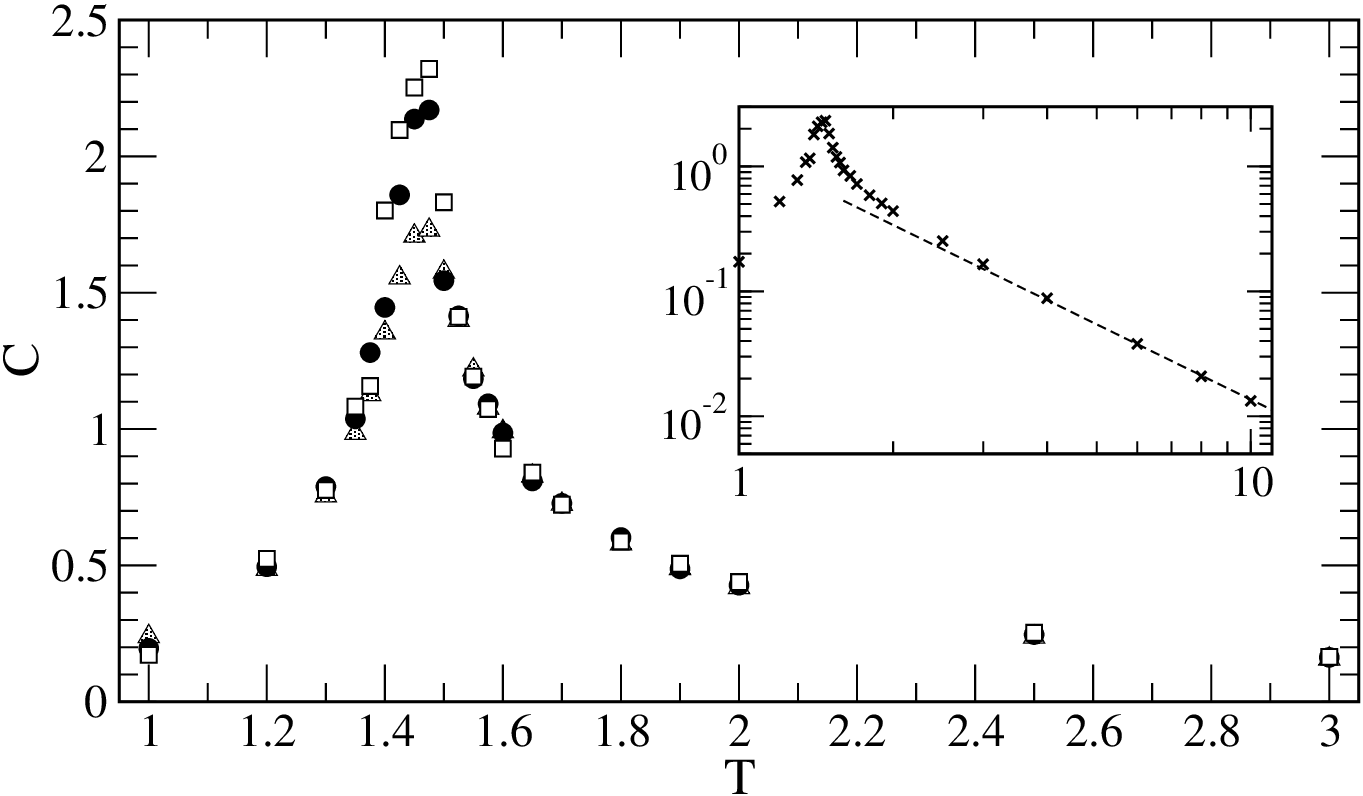}
    \caption{Temperature dependence of specific heat $C$,
      Eq.~(\protect{\ref{eq:def-C-fluct-E}}), for $L = 10$ ($\blacktriangle$),
      20 ($\bullet$) and 40 ($\square$).  The inset presents a log-log plot of
      the same quantity (for $L=40$), with the dashed line showing an
      approximation of the form $C(T) \propto 1/T^2$.
     \label{fig:heat3}
    }
\end{figure}

\begin{figure}
\includegraphics[width=\columnwidth,clip=true]{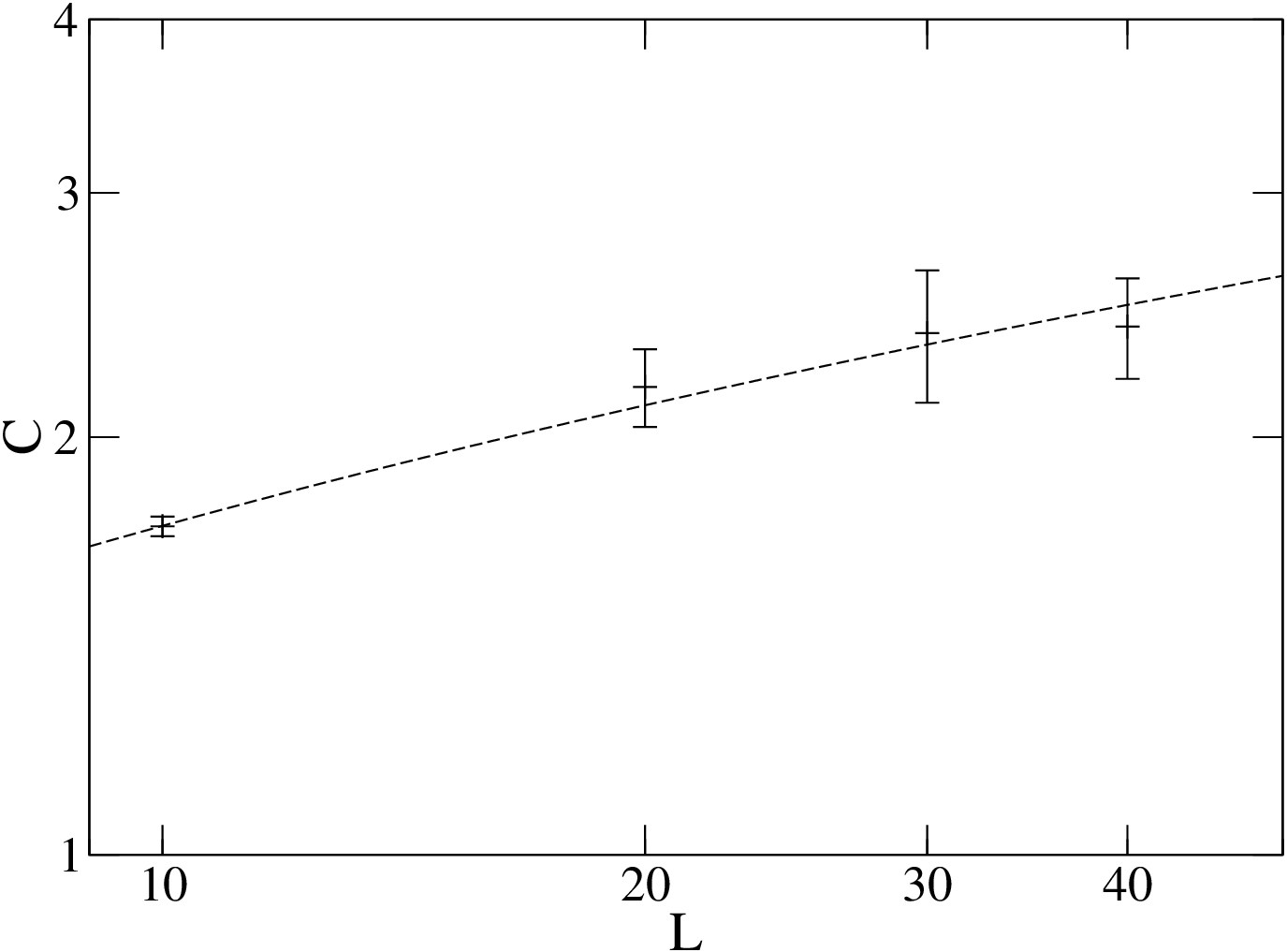}
    \caption{Finite-size scaling of the specific heat $C$ at $T=1.475$
      (log-log plot).  The dashed line is the best fit to the formula $C(L) =
      a + b\ln(L)$.
     \label{fig:cv_L}
    }
\end{figure}

The temperature dependence of the specific heat, calculated for three
different lattice sizes from energy fluctuations,
\begin{equation}
    C = \frac{N}{T^2}\left( \left< u^2\right> - \left< u\right>^2 \right),
    \label{eq:def-C-fluct-E}
\end{equation}
is depicted in Fig.~\ref{fig:heat3}. As can be seen, $C$ develops a maximum
near $T_c\approx1.475$, and this maximum grows with lattice size $L$.  {
  This $L$-dependence of $C$ on the system size is far more significant near
  $T_C$ than far from it and thus should be attributed to the classical phase
  transition rather than to the effects induced by the SOC constraints. 
  The finite-size scaling theory predicts that if there is a second-order
  transition at $T_c$, $C(T_c)$ should grow as $L^{\alpha/\nu}$, with $\alpha,
  \nu$ being the usual scaling exponents.  The dependence of $C$ at $T=1.475$
  on $L$ is shown in Figure \ref{fig:cv_L} as a log-log plot. It suggests that
  the slope tends to $0$, which implies $\alpha = 0$. This, in turn, suggests
  that the hybrid system could belong to the same universality class as the
  standard Ising model, for which $\alpha = 0$, $\nu=1$, and the specific heat
  diverges logarithmically. To verify this possibility, in Fig.\ 
  \ref{fig:cv_L} we also plotted a line which represents the best-fit of the
  data to the ansatz $C(L) \approx a + b\ln(L)$. The fit turns out to be rather
  convincing.}

Notice the inset in Fig.~(\ref{fig:heat3}). It depicts $C(T)$ on a log-log
plot.  It can be observed that for large temperatures $T$ the specific heat
decays as $T^{-2}$, as it should do for {usual} Hamiltonian lattice spin
models.

\begin{figure}
\includegraphics[width=\columnwidth,clip=true]{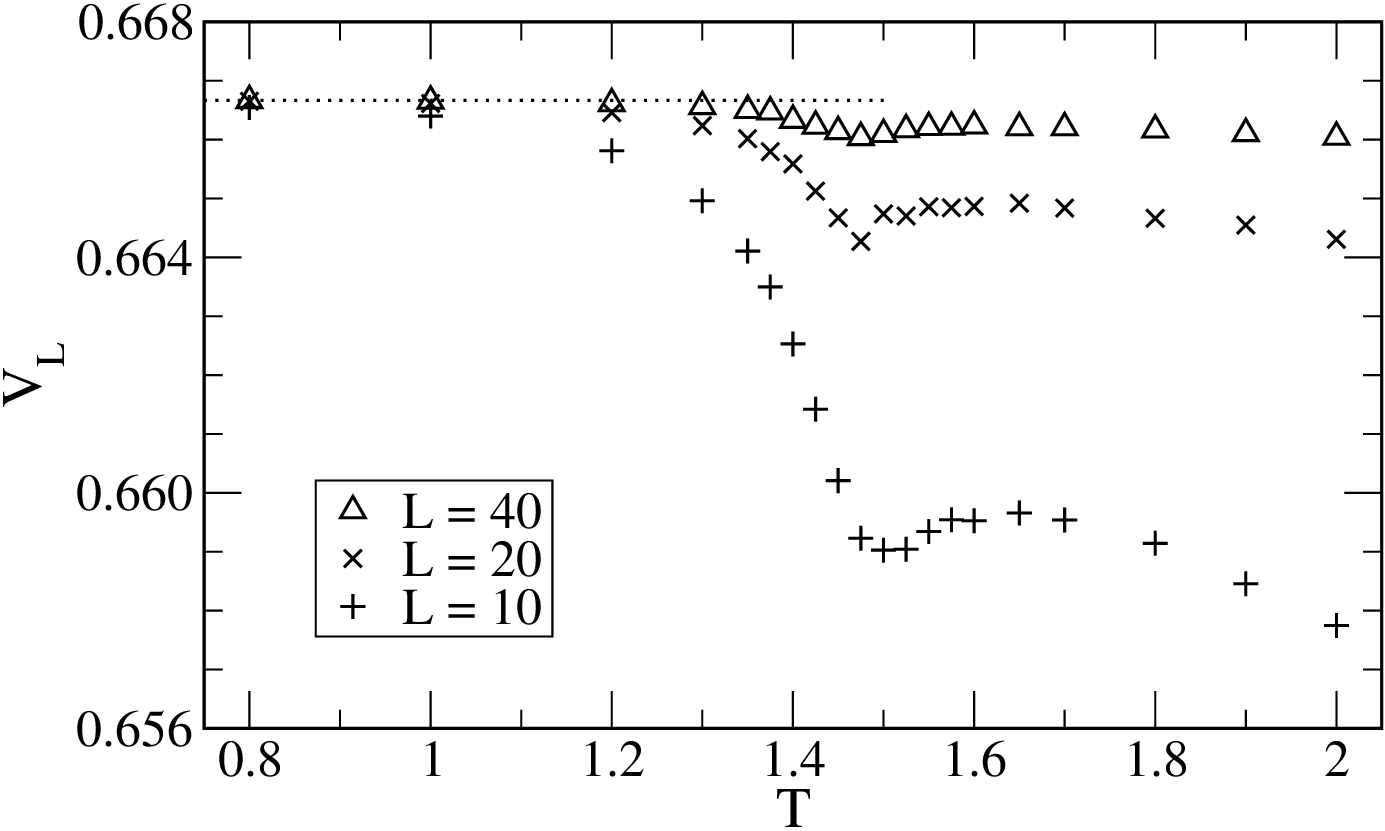}
    \caption{Temperature dependence of the energy cumulant $V_L$ for $L=10$, 20,
and 40. The dotted line shows the theoretical value of $V_L = 2/3$ as $T\to 0$
    \label{fig:fig_ce4}
    }
\end{figure}

A well-established numerical method for analysis of critical points consists in
studying the temperature dependence of the fourth cumulant of $u$,
\begin{equation}
    V_L = 1 - \frac{\left<u^4\right>}{3\left< u^2\right>^2}.
    \label{eq:Vl}
\end{equation}
This quantity is supposed to have a local minimum at $T_c$, both for continuous
and discontinuous phase transitions \cite{Binder97,Vollmayr93}. For first-order
transitions the depth of this minimum, $2/3-V_L^{\mathrm{min}}$, carries
information about the latent heat. The dependence of $V_L$ on temperature in
our model is presented in Fig.~\ref{fig:fig_ce4}. The curves develop a clear
minimum near $T=1.475$. Since its depth decreases with $L$, it suggests that
there is no latent heat and the transition is of the second order.

\begin{figure}
\includegraphics[width=\columnwidth, clip=true]{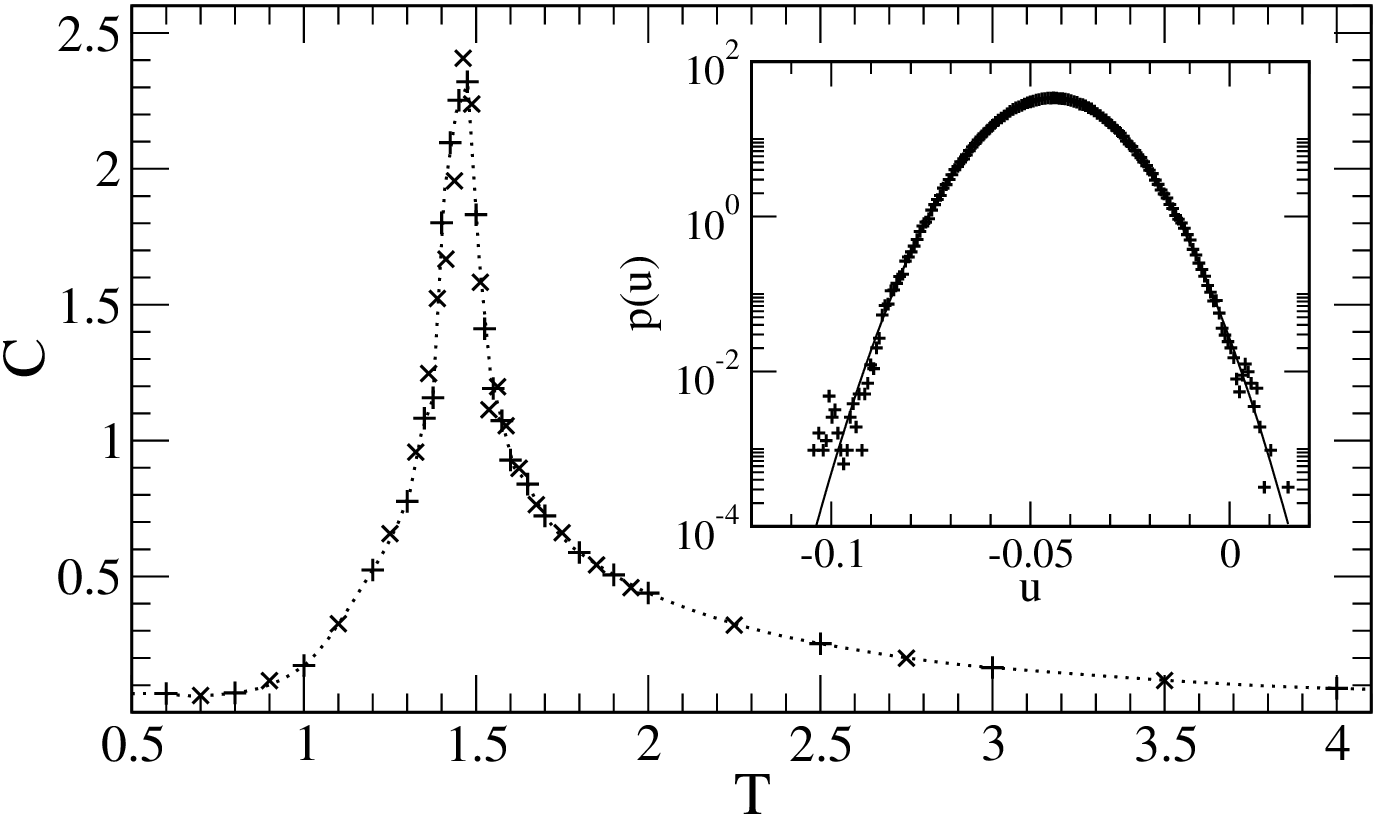}
    \caption{Specific heat per node, $C$, for $L=40$, calculated from energy
    fluctuations ($+$ and dotted line) and the derivative $du/dT$ ($\times$).
    The inset depicts the histogram of internal
    energy for $T=10$ and $L=40$.
    The solid line is the best-fit normal distribution.
    \label{fig:heat} }
\end{figure}

At this point we apply two tests on $u$ to see if our system has some basic
properties of standard models of statistical physics. First, we verify whether
the thermodynamic and microscopic definitions of specific heat,
Eqs.~(\ref{eq:C=du/dT}) and (\ref{eq:def-C-fluct-E}), are equivalent.  The
result of this test is shown in Fig.~\ref{fig:heat}, which presents $C$
calculated with the two methods. The consistency of the results is very good.
Second, we check if the distribution of energy (far from $T_c$) is normal.  An
example of a result of this test is presented in the inset of
Fig.~\ref{fig:heat}, which shows a histogram of $u$ as well as the best-fit
Gaussian approximation.  In this case the empirical and theoretical
distributions are very close to each other.


\subsubsection{The order parameter and Binder's cumulant} 
\begin{figure}
\includegraphics[width=\columnwidth]{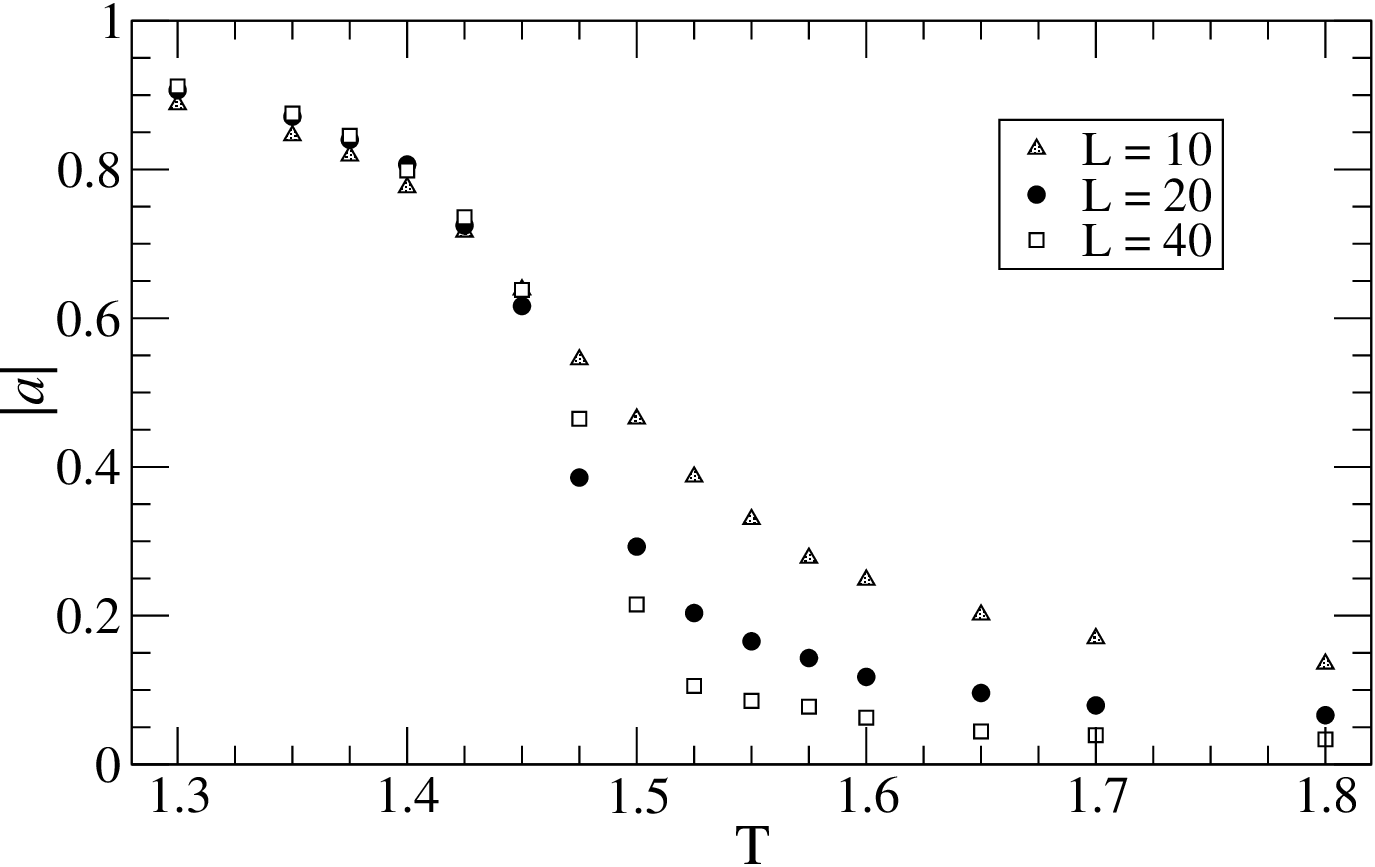}
    \caption{The antiferromagnetic order parameter $|a|$
    as a function of temperature $T$ for $L=10$, $20$ and $40$.
    \label{fig:aferro} }
\end{figure}

The temperature
dependence of the antiferromagnetic order parameter $|a|$ is depicted in
Fig.~\ref{fig:aferro}. It shows that the graphs of $|a(T)|$ for various system
sizes $L$ cross near $T= 1.475$ (probably at different temperatures).  In the
same region the slope of $|a(T)|$ gets steeper with increasing $L$, while for
higher temperatures $|a|$ quickly tends to $0$ as $L\to\infty$.  All these
properties are typical hallmarks of a {classical} phase transition.
However, large error bars (not shown) prevented us from doing any reliable
quantitative estimates of the critical exponents or $T_C$ based on these data.

\begin{figure}
\includegraphics[width=\columnwidth, clip=true]{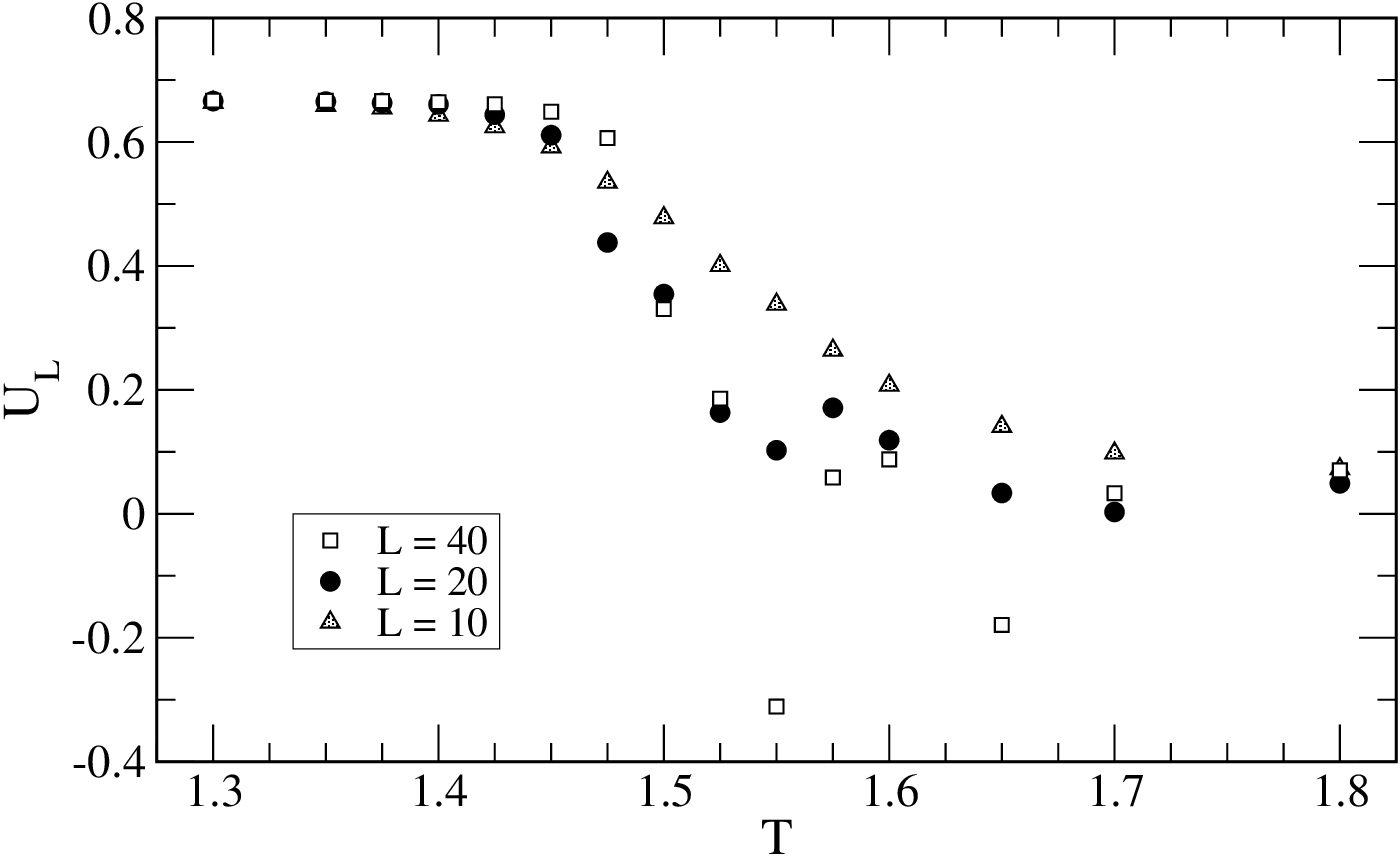}
    \caption{Binder cumulant $U_L$ as a function of temperature $T$
    for $L=10$, $20$, and $40$.
    \label{fig:Binder} }
\end{figure}

We have also performed a study of Binder's cumulant, which is the fourth
cumulant of the order parameter \cite{Binder97},
\begin{equation}
    U_L = 1 - \frac{\left< a^4 \right>}{3\left< a^2\right>^2}.
    \label{eq:Binder}
\end{equation}
This quantity is particularly useful in finding a critical temperature, $T_c$,
which can be extrapolated from the abscissas of intersection points of
$U_L(T)$ for several $L$, as these abscissas are typically only very weakly
dependent on $L$. This method is supposed to give precise results irrespective
of the order of the phase transition \cite{Binder97}.  The plot of $U_L(T)$
for our model is shown in Fig.~\ref{fig:Binder}. As can be seen, Binder's
cumulants seem to cross each other, though the cutting point coordinates are
loosely dependent on $L$. Moreover, the error bars (not shown), especially for
larger temperatures, are very high.

\begin{figure}
\includegraphics[width=\columnwidth, clip=true]{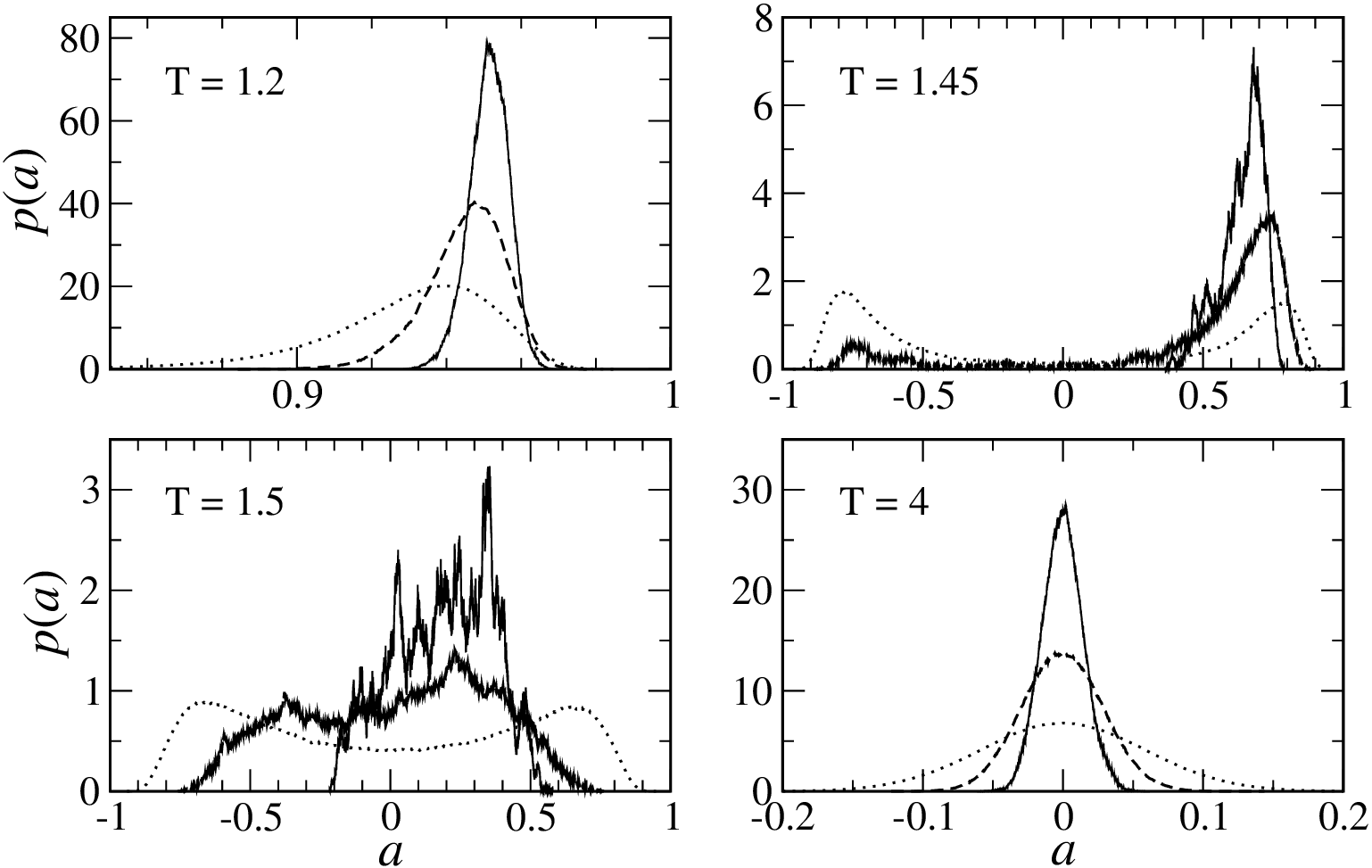}
    \caption{The histograms of the antiferromagnetic order parameter $a$ for
    $T = 1.2$, $1.45$, $1.5$, $4$ and three lattice sizes $L=10$ (dotted
    line), 20 (dashed line) and 40 (solid line, narrowest distribution).
    \label{fig:4panels} }
\end{figure}

To identify the source of these problems, we investigate histograms of the
order parameter $a$ at various temperatures. This is motivated by the fact
that  the theory behind Binder cumulant $U_L$ is based on an assumption that
these histograms can be approximated by a normal distribution in a disordered
phase, and by a superposition of two normal distributions in an ordered phase.
Four typical histograms are presented in Fig.~\ref{fig:4panels}.  The first
one shows that, for temperature $T=1.2 \ll T_c$, the histograms are smooth
functions of $a$ that can be approximated by a Gaussian only for the largest
system size ($L=40$); for small $L$ they develop a slowly decaying tail on the
left.  The second panel shows the situation for $T=1.45$, which is just below
$T_c$.  For $L=10$ the system can freely switch between states with positive
and negative $a$, but such a switch is impossible for $L=40$. In the
intermediate case of $L=20$, the system can switch to $a<0$, but the histogram
is very asymmetric, which indicates that the actual relaxation time must be
\emph{much} larger than the simulation time.  As illustrated in the third
panel, at a little higher temperature $T=1.5$ the system can already freely
switch between $a>0$ and $a<0$ for all investigated system sizes. However, the
histogram for $L=40$ turns out asymmetric about $a=0$ and very erratic, which
again indicates that the relaxation time is far larger than that available in
computer simulations. Finally, the last panel shows that at a  high
temperature, $T=4$, histograms are again smooth functions of $a$, and each can
be approximated by a normal distribution. Therefore, for temperatures
$T\approx 1.475$ and large system sizes $L$, the actual relaxation time is much larger 
than the simulation time. Large relative errors of the data presented in Fig.~\ref{fig:Binder}
can thus be explained by the phenomenon of slow relaxation, which seems to be
a generic property of sandpile models \cite{Dickman03}, and which should be
particularly important at temperatures close to $T_c$.

\subsection{Influence of the Ising interactions on criticality of the BTW
subsystem}
\label{sub:Influence}

Recall that a signature of criticality is the power-law decay
for the probability $P_{kl}(r)$ that two nodes at a distance $r$ apart have
heights $k$ and $l$ ($0 \le k,l \le 3$).  It was proved \cite{Majumdar91} that
in the bulk of a two-dimensional sandpile one has
\begin{equation}
     P_{00}(r) = P_0^2 + cr^{-4} + \ldots,
     \label{eq:Pkl}
\end{equation}
where $P_0$ and $c$ are some constants that depend on the lattice and boundary
conditions \cite{Majumdar91}. This relation can be used as a test of the
extent to which the self-organized criticality is affected by temperature and
interactions. The idea is simple: if for $T > T_c$ the system is dominated by
interactions, $P_{00}(r)$ should decay in a way typical of high-temperature
Hamiltonian systems, i.e.\ exponentially; if the system is dominated by SOC
constraints, $P_{00}(r)$ should decay in accordance with (\ref{eq:Pkl}).

\begin{figure}
\includegraphics[width=\columnwidth, clip=true]{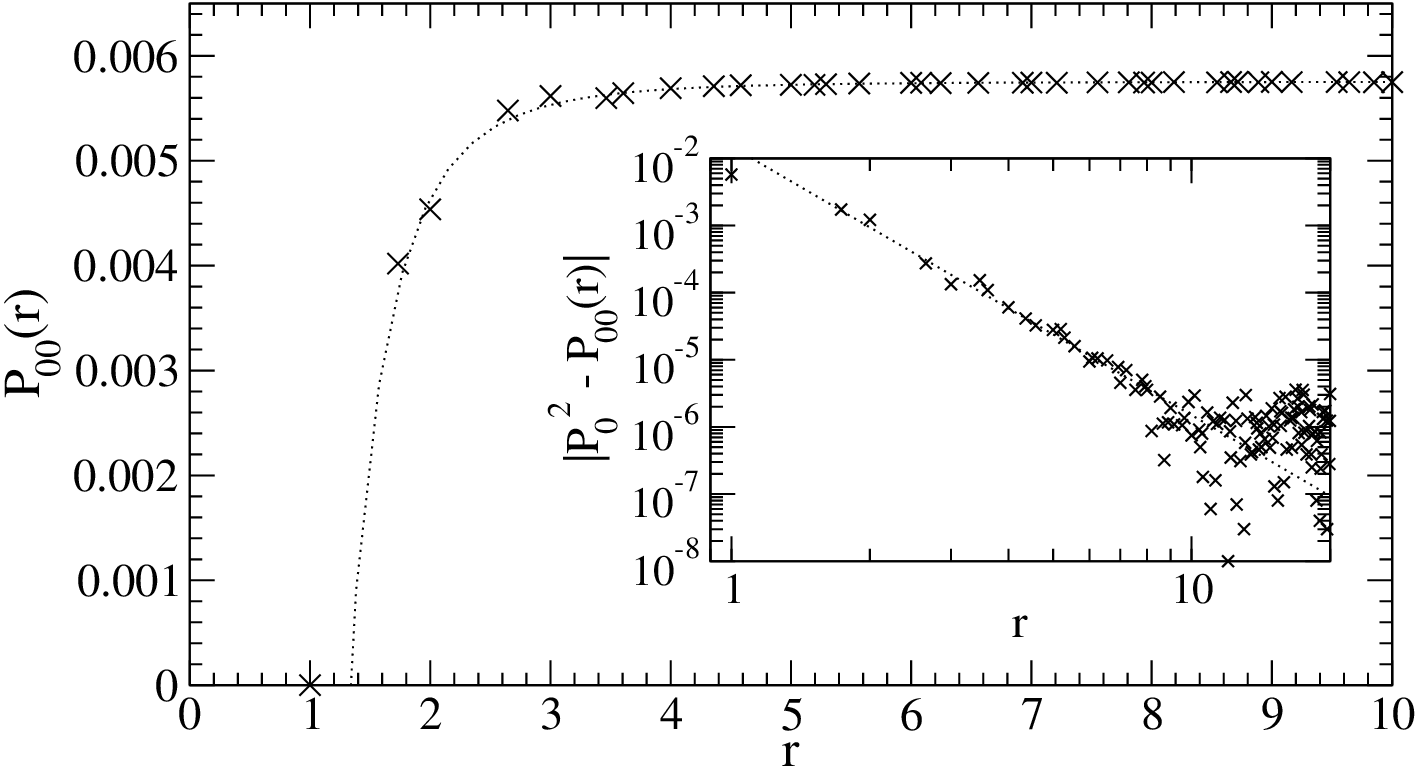}
\caption{Probability $P_{00}(r)$ of finding two empty lattice
    nodes $r$ units apart for $L=40$ and $T = 10$. The
    dotted line is a fit calculated from (\protect\ref{eq:Pkl}).
    The inset presents the log-log plot of  $|(P_0)^2 - P_{00}(r)|$.
    The error bars are of order $10^{-6}$ (not shown).
     \label{fig:fig_hcorels}
     }
\end{figure}

The results for a rather high temperature $T=10$, where the Ising interactions
should not play a significant role, are depicted in
Fig.~\ref{fig:fig_hcorels}.  The log-log plot in the inset confirms validity
of Eq.\ (\ref{eq:Pkl}).  Similar results were also obtained for all
temperatures $T \gtrsim 2.5$, whereas for $T < 2.5$ an exponential fit was
also possible. We explain this as follows.  Generally one can expect that for
temperatures $T > T_c$ the decay rate of $P_{00}(r)$ will have both power-law
(SOC) and exponential (thermodynamic) contributions, the former proportional
to $1/r^4$ and the latter proportional to $\exp(-r/\lambda)$, where $\lambda$
is the correlation length.  For $r$ large enough, the power-law should
dominate, but for smaller distances, like those available in computer
simulations, $P_{00}(r)$ may be dominated by the exponential term, especially
for temperatures closer to (but larger than) $T_c$, as in this case the
thermodynamic correlation length gets large and the exponential term can
exceed the rapidly decaying algebraic term $1/r^4$.  Another reason why we
did not confirm (\ref{eq:Pkl}) for $T < 2.5$ is that as the system gets
cooler, the Ising interactions become more pronounced. As one of their effect
is to reduce the number of sites with $h_j=0$, the statistics for exploring
$P_{00}(r)$ becomes too poor for reliable data analysis.

Validity of equation (\ref{eq:Pkl}) for $T>T_c$ can be also justified by the
following
physical argument. Suppose that the role of interactions, above $T_c$, is
restricted to flipping some spins from $S_j=0$ to $S_j=\pm1$, with a
probability $p$, in an \emph{uncorrelated} manner. This reduces $P_{00}(r)$ by
a factor $(1-p)^2$. Consequently, $P_{00}(r)$ should satisfy (\ref{eq:Pkl})
also for finite temperatures, with the ratio $P_0^2/c$ independent of $T$.
And indeed, our
estimates of this ratio for $T=2.5, 3, 4, 6, 8,$ and $10$ yield a constant value
$0.34\pm 0.01$. 

We conclude that the system preserves the SOC-induced correlations at all
temperatures $T>T_c$. Note, however, that this has absolutely no impact on the
thermodynamic properties of the system.

\section{Conclusions and outlook}
\label{sec:conclusions}

We found that the spin-1 Ising model with nonlocal constraints imposed by the
Bak-Tang-Wiesenfeld sandpile model, which is expected to exhibit some features
of both thermodynamic and self-organized criticality, behaves just like a
standard model of statistical physics, with all hallmarks of a continuous
phase transition. Our analysis shows that the power-law correlations induced
by the `self-organized criticality' of the sandpile model have no impact on
the order of the phase transition in the Hamiltonian subsystem. {In
  particular, we found that the critical exponent $\alpha\approx0$, which 
suggests that the hybrid model belongs to the same universality class as the
standard Ising model.}
Similarly, Hamiltonian interactions of the Ising model cannot destroy
``criticality'' (i.e.\ power-law correlations) of the BTW sandpile model above
the thermodynamic $T_c$.  However, unlike thermodynamic criticality,
long-range correlations induced by the self-organized criticality do not show
up in thermodynamic functions {(for example, they do not affect temperature
  dependence of the specific heat)}.  Clearly, the
long-range order induced by Hamiltonian interactions has different impact on
the thermodynamics than that induced by using the SOC model to reduce the
phase space of the system.

Even though the spin and sandpile models we chose belong to the simplest
models in their kind, their combination results in a model that seems
intractable mathematically and is very difficult to treat numerically -- hence
the rather limited system sizes used in the present study.  
It is surely of
interest to find a different combination of Hamiltonian and SOC models that
would exhibit a similar combination of thermodynamic and SOC criticality
effects, yet would be easier to deal with.

Compared with the phase space of the standard Ising model, the phase space
volume of the hybrid model is significantly reduced (from $3^N$ down to $a^N$ with $a<3$)
in a non-local manner.
Such reduction resembles  
that associated with lowering of the system dimensionality and hence may affect
the critical properties of the system. We have shown that it actually
decreases the number of possible phases. Even though our study indicates
that the critical exponent $\alpha$ in the hybrid model takes on the same
value as in the Ising model, the question remains whether the two
systems belong exactly to the same universality class and other exponents
should be also examined.  Another problem
is to investigate the decay of correlation functions as $T\downarrow T_C$.

Last, but not least, our numerical results indicate that in the BTW model on 
a honeycomb lattice of linear size $L\to\infty$, the probabilities $c^{(k)}_L$
to find
$k$ grains in a given lattice site take on particularly simple
values,
$c^{(0)}_\infty = 2/24$, $c^{(1)}_\infty = 7/24$, and  $c^{(2)}_\infty = 15/24$.

\begin{acknowledgments}
Support from Action de Recherche Concert\'ee Program of
the University of Li\`ege (ARC 02/07-293) is gratefully acknowledged.
\end{acknowledgments}


\end{document}